\documentclass[conference]{IEEEtran}
\IEEEoverridecommandlockouts
\usepackage{cite}
\usepackage{enumitem}
\usepackage{amsmath,amssymb,amsfonts}
\usepackage{algorithmic}
\usepackage{graphicx}
\usepackage{textcomp}
\usepackage{xcolor}
\usepackage{booktabs}
\def\BibTeX{{\rm B\kern-.05em{\sc i\kern-.025em b}\kern-.08em
    T\kern-.1667em\lower.7ex\hbox{E}\kern-.125emX}}
\begin{document}

\title{Extending the Frontier of ChatGPT: Code Generation and Debugging\\

}

\author{\IEEEauthorblockN{Fardin Ahsan Sakib}
\IEEEauthorblockA{\textit{Department of Computer Science} \\
\textit{George Mason University}\\
Fairfax, USA \\
fsakib@gmu.edu}
\and
\IEEEauthorblockN{Saadat Hasan Khan}
\IEEEauthorblockA{\textit{Department of Computer Science} \\
\textit{George Mason University}\\
Fairfax, USA \\
skhan225@gmu.edu}
\and
\IEEEauthorblockN{A. H. M. Rezaul Karim}
\IEEEauthorblockA{\textit{Department of Computer Science} \\
\textit{George Mason University}\\
Fairfax, USA \\
akarim9@gmu.edu}

}

\maketitle

\begin{abstract}
Large-scale language models (LLMs) have emerged as a groundbreaking innovation in the realm of question-answering and conversational agents. These models, leveraging different deep learning architectures such as Transformers, are trained on vast corpora to predict sentences based on given queries. Among these LLMs, ChatGPT, developed by OpenAI, has ushered in a new era by utilizing artificial intelligence (AI) to tackle diverse problem domains, ranging from composing essays and biographies to solving intricate mathematical integrals. The versatile applications enabled by ChatGPT offer immense value to users. However, assessing the performance of ChatGPT's output poses a challenge, particularly in scenarios where queries lack clear objective criteria for correctness. For instance, evaluating the quality of generated essays becomes arduous and relies heavily on manual labor, in stark contrast to evaluating solutions to well-defined, closed-ended questions such as mathematical problems. This research paper delves into the efficacy of ChatGPT in solving programming problems, examining both the correctness and the efficiency of its solution in terms of time and memory complexity. The research reveals a commendable overall success rate of 71.875\%, denoting the proportion of problems for which ChatGPT was able to provide correct solutions that successfully satisfied all the test cases present in Leetcode. It exhibits strengths in structured problems and shows a linear correlation between its success rate and problem acceptance rates. However, it struggles to improve solutions based on feedback, pointing to potential shortcomings in debugging tasks. These findings provide a compact yet insightful glimpse into ChatGPT's capabilities and areas for improvement.
\end{abstract}

\begin{IEEEkeywords}
ChatGPT, Code Generation, Programming Problems, Debugging
\end{IEEEkeywords}

\section{Introduction}
\noindent Artificial intelligence has achieved remarkable prowess across a plethora of dimensions, encompassing code generation \cite{tian, bavishi, li, siddiq, svyatkovskiy}, program explanation \cite{hu,lib,stapleton,wang}, error correction \cite{gupta,jiang,dear,yeh}, and more. Within the domain of code generation, these AI tools exhibit a remarkable capability to write programs based on natural language descriptions of a given problem. Recent years have witnessed an unprecedented surge in the development of neural network architectures, notably the transformer model\cite{attention}. This surge has started a new era of large-scale language models (LLMs), empowering users with the ability to leverage pre-trained models that have been extensively trained on vast amounts of code and natural language data.\cite{chen, bert,tianh,zhang,lertbanjongngam}\\
The advent of these large language models has sparked a cosmic shift in the landscape of code generation. With their immense knowledge and linguistic expertise, these LLMs have transformed the way coding problems are approached. By bridging the gap between human language and programming language, these models have opened up new horizons for programmers and researchers alike. \\
OpenAI's\cite{openai} ChatGPT\cite{chatgpt} exemplifies the epitome of AI tools, harnessing the power of large language models (LLMs) integrated within a user interface to engage in interactive, conversational exchanges and generate responses. This tool transcends boundaries, catering to an extensive array of user needs expressed in natural language. Within the domain of code generation, ChatGPT boasts exceptional features that enable it to answer queries pertaining to programming challenges while maintaining logical coherence. These distinctive attributes of ChatGPT can be listed as such: 
\begin{itemize}
    \item \textbf{Knowledge and Pattern Recognition: }By virtue of its extensive training on a vast dataset encompassing software development and programming languages, the model has gained a deep understanding of the essence of programming problems, discerning intricate patterns within inputs, and expertly leveraging its vast knowledge to generate accurate and contextually appropriate solutions. 
    \item \textbf{Natural Language Processing Capability: }At the core of ChatGPT's capabilities lies its aptitude in natural language processing (NLP), endowing it with the ability to comprehend and interpret human language. The neural network model powering ChatGPT has been meticulously trained to grasp the nuances of human expression, enabling it to generate outputs that effectively address queries posed in natural language.
    \item \textbf{Generalizing Capability: }Impressively, ChatGPT possesses a remarkable generalization capability, transcending the confines of its training data. Although trained on an extensive dataset comprising a vast corpus of code, the model demonstrates exceptional performance when confronted with novel queries that lie beyond the scope of its training. This generalization capability further enhances its utility and flexibility, making it an invaluable tool for developers and programmers seeking comprehensive code generation support.
\end{itemize}
Amidst the rising popularity of ChatGPT among programmers, it is imperative to acknowledge that its responses to programming problems expressed in natural language may exhibit imperfections. Consequently, a comprehensive study is imperative to fully assess ChatGPT's effectiveness in addressing such problems. This research endeavors to conduct a meticulous evaluation of ChatGPT's code generation capabilities, as well as its debugging aptitude. A central focus is placed on assessing the model's performance across a diverse spectrum of programming problems, spanning various domains and levels of complexity. To facilitate this study, a carefully curated custom dataset comprising programming problems from Leetcode\cite{leetcode} is employed, encompassing a rich variety of problem scenarios.\\
In addition to evaluating ChatGPT's code generation, the research uniquely delves into its debugging capabilities. After ChatGPT initially fails to produce correct solutions for certain problems, Leetcode's feedback and error messages are provided to the model in an attempt to prompt improvements in its solutions. The research meticulously analyzes ChatGPT's response to this feedback, elucidating its capacity to learn from errors and rectify its solutions. By incorporating this evaluation of ChatGPT's debugging aptitude, the research offers a holistic assessment of the model's performance, addressing both its strengths and limitations in solving programming problems.\\
Upon concluding the study, several key findings have emerged:
\begin{itemize}
    \item ChatGPT showcased a strong overall performance, by providing correct solution to 71.875\% of the problems present in the constructed dataset. Notably, it exhibited particular proficiency in structured domains such as ``Tree'' and ``Divide and Conquer''. However, it encountered challenges when confronted with complex problems falling under ``Greedy'' and ``Dynamic Programming (DP)'' domains.
    \item The model's success rate displayed a linear correlation with the acceptance rates of the problems. It excelled in problems with higher acceptance rates while facing difficulties in problems with lower acceptance rates.
    \item ChatGPT demonstrated limited adaptability in response to feedback, improving its solutions in only 36.7\% of cases. This finding suggests potential weaknesses in the model's debugging and error-learning capabilities, which refers to its ability to learn from the feedback provided, analyzing the errors in the initially generated solutions, which seems to be constrained, warranting further research and enhancement to bolster its capacity in effectively incorporating feedback for improved performance.
\end{itemize}
These findings provide valuable insights into ChatGPT's performance in solving programming problems, allowing users to make informed decisions about utilizing the tool and emphasizing areas for future enhancement and refinement.\\

\section{Literature Review}
\begin{figure}[b]
\centerline{\includegraphics[width=0.45\textwidth]{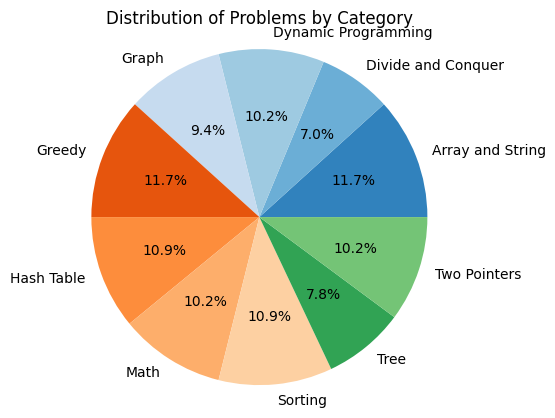}}
\caption{The dataset exhibits a balanced distribution across problem domains, with approximately equal proportions of questions allocated to each category.}
\label{problem-domain}
\end{figure}
\noindent In recent years, large language models (LLMs) have significantly influenced the field of natural language processing. These models, trained on vast text and code datasets, exhibit exceptional performance on a wide range of tasks including text generation, code generation, machine translation, question-answering, summarization, and more\cite{nlpmt, nlpqa, nlpcodegeneration, nlpsummarization}. Such advancements have far-reaching impacts across several domains such as education, healthcare, finance, and customer service \cite{education, biomedical, codegeneration,finqa, tnrflood }.\\
GPT-1 \cite{GPT1}, a pioneering LLM, was trained on the task of next-word prediction, enabling it to understand word dependencies and generate relevant, context-aware content. Its successor, GPT-2 \cite{GPT2}, was trained on a much larger dataset and incorporated 1.5 billion parameters, significantly improving its performance in a zero-shot learning setting. GPT-3 \cite{GPT3} marked a considerable advancement, boasting an impressive 175 billion parameters.\\
ChatGPT, a conversational AI model developed by OpenAI, is currently considered one of the most advanced LLMs due to its sophisticated training techniques and large training corpus. Notably, it is powered by GPT-4 \cite{gpt-4}, the latest iteration of the model.\\
The concept of using these LLMs for automatic code generation has been a significant area of research, given its potential to reduce human error and boost efficiency.\\
For example, Dakhel et al. \cite{dakhel} assessed the efficacy of GitHub Copilot in generating solutions for fundamental algorithmic problems. While the model was able to provide correct solutions for several problems, it fell short of matching human programmers' performance. Prenner et al. \cite{prenner} explored Codex, a pre-trained LLM, and its ability to identify and rectify bugs in code. Their research showed that Codex is extraordinarily effective, even competitive with leading automated program repair techniques, especially in fixing Python bugs. Sobania et al.\cite{sobania} evaluated ChatGPT's capability in code repair using the QuixBugs benchmark set. By harnessing its conversational abilities, ChatGPT successfully repaired 31 out of 40 bugs, outperforming established methods. Xia et al. \cite{xia} presented the enhanced performance of conversational automated program repair (APR) over earlier LLM-based APR approaches. Chen et al. \cite{chen} developed Codet, a tool that uses a pre-trained language model to generate test cases for code samples, a feature that can significantly improve code quality and correctness.\\
Tian et al. \cite{tian} investigated ChatGPT's potential as a programming assistant. Their research indicated that ChatGPT could generate correct code for a variety of problem types and difficulty levels based on the LeetCode benchmark. However, they also noted that ChatGPT struggles to generalize its code generation capabilities to unseen and novel problems.\\

\section{Research Methodology}
\subsection{Preparation of Dataset}

\begin{figure}[t]
\centerline{\includegraphics[width=0.45\textwidth]{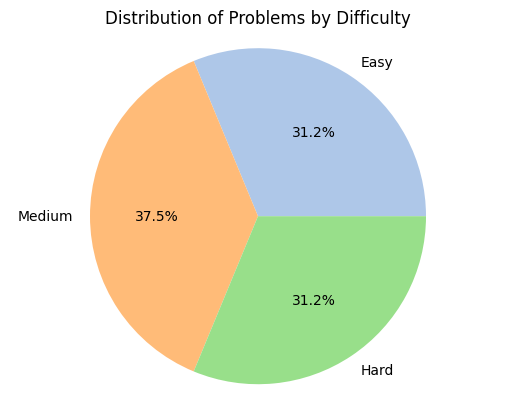}}
\caption{With nearly equal proportions of questions allocated to each respective difficulty level, the dataset showcases a well-balanced distribution.}
\label{fig2}
\end{figure}
\noindent In order to ascertain the practical applicability of ChatGPT in addressing coding problems, a dataset was meticulously curated comprising coding challenges sourced from Leetcode, a renowned platform known for its extensive collection of coding problems. Leetcode provides users with a comprehensive array of coding questions, complete with visualizations and an integrated development environment (IDE) that facilitates coding solutions. The dataset construction process commenced with the extraction of coding questions from diverse domains that represent the most prominent areas of coding expertise, including but not limited to Hash-tables, Divide and Conquer, Greedy approach, and Graph. To ensure a comprehensive evaluation, the incorporation of coding challenges of varying difficulty levels, ranging from rudimentary problems to advanced and intricate ones was ensured. Figures \ref{problem-domain} and \ref{fig2} provide a visual depiction of the domains of problems covered in the dataset along with the difficulty variation. It is imperative to acknowledge that the data collection occurred prior to May 2023, and the outcomes presented in this study are contingent upon the specific dataset utilized. It is essential to recognize that these results may be subject to modifications if more recent data is gathered.

\subsection{Analysis of Dataset}


\begin{table*}[ht]
\caption{A quantitative summary of the dataset that presents the distribution of instances across problem domains and difficulty levels}
\centering
\renewcommand{\arraystretch}{1.5} 
\begin{tabular}{|c|c|c|c|c|c|c|c|c|c|c|c|c|c|c|c|}
\hline
\textbf{Difficulty} & \textbf{Acceptance} & \multicolumn{10}{|c|}{\textbf{Number of Problems in each category}} \\
\cline{3-12} 
\textbf{Level} & \textbf{Rate} & \textbf{\textit{Array \& String}}& \textbf{\textit{DP}}& \textbf{\textit{Divide \& Conquer}} & \textbf{\textit{Graph}}& \textbf{\textit{Greedy}}& \textbf{\textit{Hash Table}}& \textbf{\textit{Math}}& \textbf{\textit{Sorting}}& \textbf{\textit{Tree}} & \textbf{\textit{Two Pointers}}\\
\hline
\textbf{} & \textbf{$<30\%$} & 2&0 &0 & 0& 2& 2& 2& 0&0 &0 \\
\cline{2-12}
\textbf{Easy} & \textbf{$30-70\%$} &1 &1 &1 &2 & 1& 1& 1&2 &1 &3\\
\cline{2-12}
\textbf{} & \textbf{$>70\%$} & 2&2 &1 &1 &2 & 2&2 & 2&2 &2 \\
\hline 
\textbf{} & \textbf{$<30\%$} &2 &2 &1 & 2& 2&2 &2 & 2& 0& 2\\
\cline{2-12}
\textbf{Medium} & \textbf{$30-70\%$} &1 & 1&1 & 1& 1&1 & 1&1 & 2&1\\
\cline{2-12}
\textbf{} & \textbf{$>70\%$} &2 &2 &2 & 2&2 &2 & 2& 2&2 & 2\\
\hline
\textbf{} & \textbf{$<30\%$} & 2&2 &2 &2 &2 & 2& 2&2 & 0& 2\\
\cline{2-12}
\textbf{Hard} & \textbf{$30-70\%$} &1 &1 & 1&2 &1 &2 &1 & 1& 2&1\\
\cline{2-12}
\textbf{} & \textbf{$>70\%$} &2 &2 & 0& 0& 2& 0& 0&2 &1 & 0\\
\hline
\multicolumn{2}{|c|}{\textbf{Total Number of Problems}} &15&13&9&12&15&14&13&14&10&13\\
\hline
\end{tabular}
\label{table1}
\end{table*}


\begin{figure}[b]
\centerline{\includegraphics[width=0.45\textwidth]{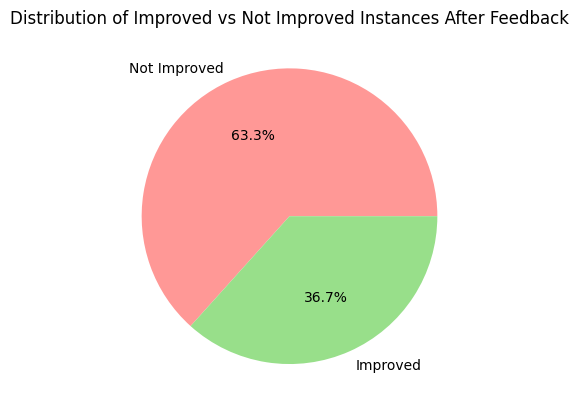}}
\caption{After receiving feedback, the model could only solve 36.7\% of the problems on a second attempt that it had previously failed on the first try, highlighting its inadequate debugging proficiency.}
\label{improved-not}
\end{figure}
\noindent The dataset construction process meticulously considered the need for diversity across various problem domains, while also accounting for the difficulty levels of the challenges and their solution acceptance rates on the Leetcode platform. The resulting dataset comprises carefully chosen problems that reflect a wide range of genres, ensuring comprehensive coverage of the coding landscape. Table \ref{table1} presents a detailed overview of the dataset, presenting the genres of problems alongside the number of problems from each difficulty level and their respective solution acceptance rates. The dataset was thoughtfully crafted to include precisely 15 problems from each problem category. Within each difficulty level (Easy, Medium, and Hard), we ensured the inclusion of five problems, following a specific acceptance rate criterion. Two problems were selected with an acceptance rate below 30\%, one problem with an acceptance rate between 30\% and 70\%, and the remaining two problems with an acceptance rate exceeding 70\%. It is essential to acknowledge that due to the scarcity of problems meeting these specified criteria in certain categories, the total number of problems in the dataset amounts to 128. Nevertheless, the careful selection process and adherence to acceptance rate distributions ensure a balanced representation of difficulty levels and problem acceptance rates within the dataset which can be noticed in the figures \ref{problem-domain} and \ref{fig2}.
\subsection{Approaching each problem}\label{AA}
\noindent After constructing the dataset, each individual problem, along with its corresponding code structure available in Leetcode, was presented as input to the ChatGPT model. The objective was to leverage the power of the large language model (LLM) to generate code solutions based on the provided code structure. The ChatGPT model, prompted with the problem description and code structure, responded by generating a code solution, which was then utilized as the suggested solution within Leetcode's integrated development environment (IDE).\\
The generated solution provided by ChatGPT after running in the IDE of Leetcode is submitted for evaluation which results in one of the following:
\begin{enumerate}[label=\arabic*.]
    \item Upon successful execution of the generated code solution, the Leetcode platform displays vital performance metrics and the solution's comparative efficiency in terms of outperforming other submitted solutions. In that case, the following steps are taken:
    \begin{enumerate}[label=\alph*.]
        \item The solution provided by ChatGPT is listed as a ``Passed Instance".
        \item The runtime of the generated solution in millisecond (ms) and the memory consumption of the executed solution in megabytes (MB) is noted.
        \item The percentage of other submitted solutions for that problem in Leetcode that this solution beats in runtime is noted.
        \item The percentage of other submitted solutions for that problem in Leetcode that this solution beats in memory consumption is noted. 
    \end{enumerate}
    \item In the event that the generated solution is not accepted by the Leetcode platform, it can be attributed to one of the following scenarios:
    \begin{enumerate}[label=\alph*.]
        \item The occurrence of a runtime error (RTE), indicating that the program encountered an error during the execution of the provided test cases.
        \item A time limit exceeded error (TLE), signifying that the program surpassed the allotted execution time designated by Leetcode.
        \item A memory limit exceed error (MLE), denoting that the program surpassed the allocated memory usage threshold specified by Leetcode.
        \item Failure to pass all the test cases provided by Leetcode, indicating that the solution does not attain complete correctness and accuracy.
    \end{enumerate}
    \item In the event of a failed solution, the error messages generated by the Leetcode platform or the failed test cases are utilized as feedback to the ChatGPT model. The model is then prompted to rectify and enhance the provided solution, effectively evaluating the model's debugging capabilities. The resulting modified solution is subsequently re-submitted to the Leetcode IDE for evaluation. 
    \begin{enumerate}[label=\alph*.]
        \item If the modified solution successfully passes all the test cases, the process proceeds to Step 1 for further analysis and assessment.
        \item Conversely, if the modified solution fails to meet the desired requirements and triggers any of the error messages encountered in Step 2, the corresponding problem is deemed as a failed attempt by ChatGPT.
    \end{enumerate}
    

\end{enumerate}
\section{Analysis of ChatGPT's Performance}
 Following the construction of the comprehensive dataset, comprising a total of 128 problems from various categories of problems mentioned earlier, we proceed to evaluate the performance of ChatGPT in generating the solutions to the programming challenges from natural language problem descriptions. The results obtained provided valuable statistical insights into the model's capabilities. 
\subsection*{\textbf{Overall Performance}}
\noindent ChatGPT exhibited an overall success rate of \textbf{71.875\%} across the entire dataset. This indicates that out of 128 problems, ChatGPT successfully generated solutions for 92 of them. Notably, among these successful cases, 84 problems were solved in the initial attempt, rest 8 were solved after prompting the ChatGPT to debug the previously provided solution with the feedback received from Leetcode. \\
There were 36 problems within the dataset for which ChatGPT did not produce satisfactory solutions, even after revisiting the problems with feedback from Leetcode.
\subsection*{\textbf{Feedback and Debugging}}\noindent Out of the 36 problems for which ChatGPT initially failed to produce any correct solution, the Leetcode platform provided error messages and feedback as guidance for improvement. This feedback, accompanied by a few of the test cases that the solution failed to pass, was utilized to prompt ChatGPT to rectify its errors and generate corrected solutions. Surprisingly, despite the availability of this feedback, ChatGPT failed to produce correct solutions in the majority of cases.\\
In fact, when ChatGPT attempted to fix its errors using the provided feedback, the new solutions exhibited a downgrade in performance. It can be seen from Figure \ref{improved-not} that approximately 63\% of the time, these revised solutions failed to pass test cases that were previously passed, indicating a decrease in solution correctness. Only around 36\% of the instances did the new solutions perform better, passing more test cases than before but yet not completely accurate.\\
The inability of ChatGPT to generate correct solutions even after receiving feedback highlights its limitations in effectively incorporating debugging information from Leetcode. This indicates a weakness in ChatGPT's current debugging capabilities, impeding its ability to learn from and rectify errors based on provided feedback. Consequently, the model's performance in improving solution correctness falls short, as demonstrated by the downgrade in performance observed in the revised solutions.
\subsection*{\textbf{Across Different Domains}} 
\begin{figure}[h]
\centerline{\includegraphics[width=0.45\textwidth]{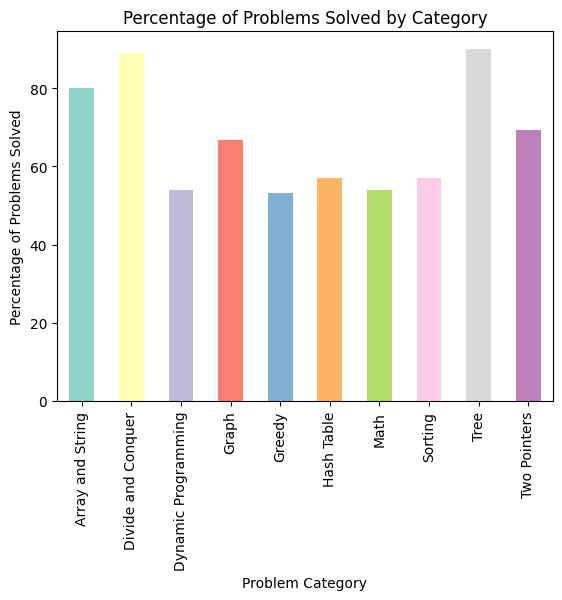}}
\caption{The comparative performance of ChatGPT across various problem domains, as indicated by pass rates, reveals that the model excels most in addressing 'Divide and Conquer' and 'Tree' problems, while demonstrating the least proficiency in tackling 'Greedy' and 'Dynamic Programming (DP)' problems.}
\label{passed-category}
\end{figure}

\noindent Analyzing the success rate of ChatGPT at solving problems across different domains reveals intriguing findings, as depicted in Figure \ref{passed-category}. The model showcases the highest success rates on problems stemming from ``Tree'' and ``Divide and Conquer'' domains. On the contrary, the model showed subpar performance while solving problems from ``Greedy" and ``Dynamic Programming (DP)" domains. \\
This observation suggests that ChatGPT has a higher success rate at solving problems that adhere to well-defined rules and structured patterns as they are easier for the model to understand and generate solutions for. Therefore, intuitively, the model finds it more challenging to solve problems that require deeper analyses and do not conform to specific rules for generating solutions that eventually lead to lower success rates. For instance, in a tree problem, it is possible for the model to easily identify the root node and then recursively solve the problem for each child node. Similarly, in a divide and conquer problem, the model can easily divide the problem into smaller subproblems, solve each subproblem, and then combine the solutions to solve the original problem.\\
In contrast, the fundamental nature of Greedy and Dynamic Programming is much different. Greedy problems start with local optimal decisions with the goal to reach an optimal global decision via the steps it takes. Similarly, in a Dynamic Programming problem, the solution comprises a sequence of decisions where it breaks the original problem into overlapping subproblems and then reuses the subproblems' solutions to solve the original problem. Such categories of problems require advanced reasoning and exploration of a wide-ranging set of possibilities that are difficult for ChatGPT to accommodate during the generation of solutions, eventually resulting in a lower comparative success rate.
These findings suggest that ChatGPT excels in generating solutions for problems that follow a well-structured methodology or adhere to established patterns. However, the model faces challenges in generating solutions for problems that require more complex decision-making processes and the analysis of various test cases. In such cases, ChatGPT's performance falls short, indicating the limitations of the model in handling problems that demand nuanced reasoning and consideration of diverse scenarios.



\begin{figure}[t]
\centerline{\includegraphics[width=0.45\textwidth]{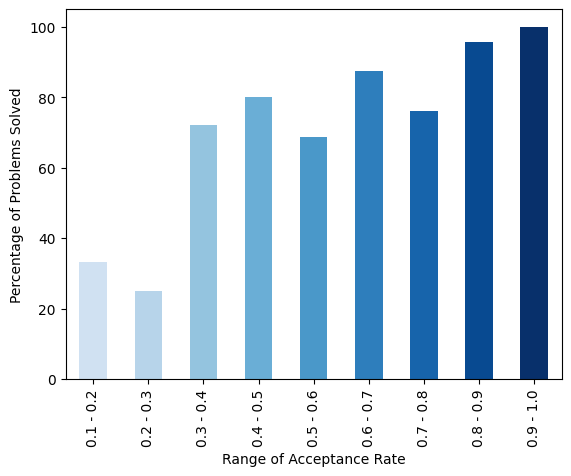}}
\caption{Analysing the percentage of problems solved with respect to the acceptance rate demonstrates a positive correlation between the acceptance rate of problems and the percentage of successful solutions generated by ChatGPT. Higher acceptance rates correspond to a higher degree of success in solving problems.}
\label{acceptance-percent}
\end{figure}
\begin{figure}[t]
\centerline{\includegraphics[width=0.45\textwidth]{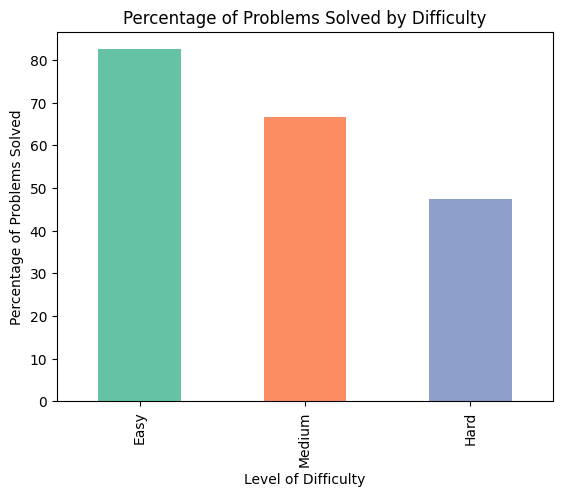}}
\caption{An analysis of pass rates at varying difficulty levels reveals that ChatGPT demonstrates a high degree of success when tasked with solving problems categorized as 'easy'.}
\label{difficulty-percent}
\end{figure}
\subsection*{\textbf{Across Different Difficulty levels and Acceptance Rates }}\noindent A similar problem trend can be observed from Figure \ref{difficulty-percent} and Figure \ref{acceptance-percent}, which provide insights into ChatGPT's success rate in solving problems of varying difficulties and acceptance rates. Notably, the model encountered challenges when tackling problems labeled as ``Hard", achieving a success rate of 55\% in generating accurate solutions. Conversely, when confronted with problems labeled as ``Easy", ChatGPT demonstrated a notably high success rate of 90\%. This observation reveals a downward trend in the percentage of successfully solved problems as the difficulty level increases. This phenomenon can be attributed to the fact that ``Easy" problems often involve well-established techniques and require less analytical prowess, while more challenging problems necessitate deeper analysis and novel problem-solving approaches, which may exceed ChatGPT's current capabilities.\\
Additionally, ChatGPT's performance is influenced by the acceptance rate of the problems. The acceptance rate reflects the success rate of all submitted solutions for a particular problem on the Leetcode platform. It is noteworthy that the dataset used in this study does not encompass problems with acceptance rates below 10\% or above 90\%. When analyzing problems with lower acceptance rates, particularly below 30\%, ChatGPT struggled to produce accurate solutions, resulting in success rates ranging from 30\% to 40\%. Conversely, as the acceptance rate increased, ChatGPT exhibited higher success rates, achieving an impressive 95.65\% success rate when the acceptance rate exceeded 80\%. \\
These observations that have been made from our experiments indicate that ChatGPT thrives in generating solutions for problems that conform to well-structured methodologies or established patterns, like Tree and Divide and Conquer problems. However, when confronted with challenges on domains whose solutions are not structured or do not have a blatant pattern for approaching the problem and require intricate test case analysis and consideration of input correlations, such as Dynamic Programming and Greedy problems, the model encounters difficulties in generating successful solutions.
\subsection*{\textbf{Run-time and Memory Efficiency }}
\begin{figure}[t]
\centerline{\includegraphics[width=0.45\textwidth]{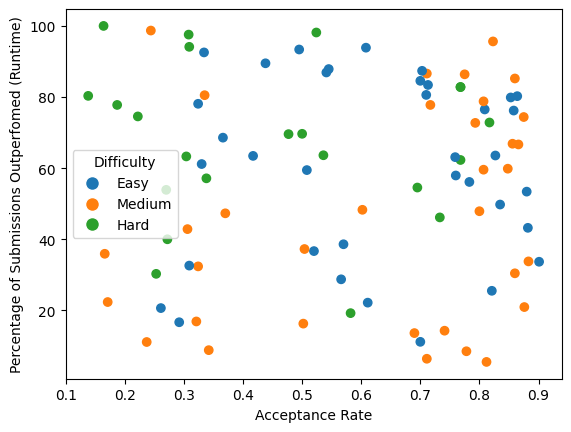}}
\caption{Scatter plot of runtime ranking against acceptance rate categorized by difficulty level}
\label{scatter-plot-runtime}
\end{figure}

\begin{figure}[t]
\centerline{\includegraphics[width=0.45\textwidth]{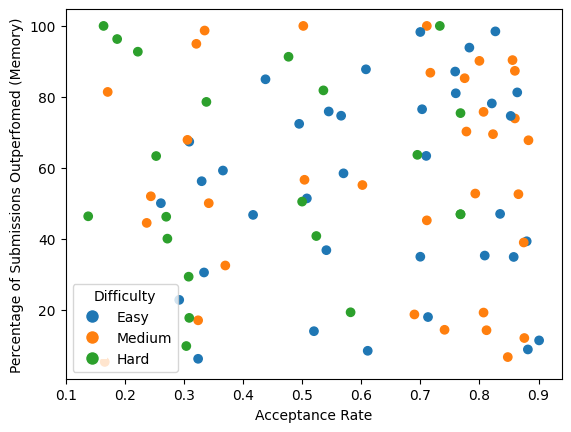 }}
\caption{Scatter plot of memory ranking against acceptance rate categorized by difficulty level}
\label{scatter-plot-memory}
\end{figure}

\noindent Let us shift our focus from the broader dataset to a more specific subset comprising instances where ChatGPT successfully solved problems. Figure \ref{scatter-plot-runtime} provides a visual representation of the runtime distribution of these successful instances, categorized by the difficulty level of the problems and the acceptance rate of the solutions. The runtime ranking metric indicates the percentage of submitted solutions on Leetcode that ChatGPT outperformed in terms of runtime efficiency. Similarly, Figure \ref{scatter-plot-memory} illustrates the percentage of submitted solutions on Leetcode that ChatGPT's solution surpassed in terms of memory efficiency. These analyses offer valuable insights into ChatGPT's ability to optimize runtime and memory utilization while generating solutions for the presented problems.\\
\noindent Upon analyzing Figure \ref{scatter-plot-runtime} and \ref{scatter-plot-memory}, several notable trends emerge, providing insightful observations regarding the efficiency of ChatGPT in terms of solution run-time and memory efficiency. Firstly, there is a discernible correlation between the difficulty level of ``Easy'' problems and their higher acceptance rates, resulting in higher efficiency in terms of run-time and memory usage. This means that ChatGPT was able to develop solutions that are efficient in terms of running time and memory utilization for these problems.\\
However, no distinct trend is apparent when examining problems classified as ``Medium'' difficulty level. This suggests that the relationship between problem difficulty and run-time or memory efficiency may not be as straightforward in this situation.\\
Remarkably, an intriguing phenomenon is observed among problems categorized as ``Hard'' with lower acceptance rates. Contrary to initial expectations, these problems exhibit higher efficiency in terms of run-time as well as memory consumption for ChatGPT's solutions. These findings run counter to intuition, as problems marked as ``Hard'' with lower acceptance rates are anticipated to pose greater challenges for the model.\\
The efficiency patterns of ChatGPT in terms of run-time and memory usage reveal complex relationships between problem difficulty, acceptance rate, and the model's performance.\\


\section{Conclusion and Future Direction}
\noindent ChatGPT represents a groundbreaking leap in the realm of AI-driven code generation, showcasing a remarkable success rate across a diverse array of coding problems. Its proficiency in effectively tackling structured problem domains and the linear correlation between its success rate and problem acceptance rates underscore its capabilities. However, it is important to acknowledge that while ChatGPT excels in generating solutions, it does not consistently produce the most efficient solutions in terms of runtime or memory usage. Additionally, approximately 30\% of the time, it generates inaccurate solutions that cannot be rectified even with feedback provided by Leetcode. This highlights the limitations of the model in certain problem scenarios.\\
Nevertheless, this exploration into the capabilities of ChatGPT underscores its potential to revolutionize code generation and assist programmers in their tasks. It serves as a testament to the power of AI-driven tools in augmenting coding workflows. Furthermore, it emphasizes the necessity for continuous refinement and evolution to enhance the capabilities of LLMs like ChatGPT, paving the way for more efficient, intuitive, and powerful coding assistance in the future.
%
%





\vspace{12pt}

\end{document}